# Trace anomaly effective actions - a critique

James M. Bardeen

*Physics Department, Box 1560, University of Washington*

*Seattle, Washington 98195-1560. USA*

*bardeen@uw.edu*

**Abstract**

The existence of trace anomalies for conformally coupled quantum fields in curved spacetimes, despite the vanishing of the trace of the conformally invariant classical stress-energy tensor, was firmly established in the 1970's, and expressions for the anomalous trace of the semi-classical quantum stress-energy tensor constructed from the curvature tensors of the background spacetime were proven to hold independent of the quantum state of the field. Subsequently, effective actions were devised whose variation with respect to the metric gives a conserved "anomalous stress-energy tensor" or ASET. These ASETs, sometimes together with a contribution from a conformally invariant contribution to the effective action, were used to derive approximations to the full semi-classical stress-energy tensor (SCSET), which is state dependent. I argue that the ASET should be defined as state independent, and for a Schwarzschild black hole spacetime should be regular on the horizon and should fall off at infinity fast enough to be consistent with asymptotic flatness. Previously proposed ASETs are unsatisfactory in this regard, but I show that it is possible to meet the above criteria. This ASET is evaluated and compared with numerical results for the full SCSET.

## I. INTRODUCTION

A conformally coupled classical field has a traceless stress-energy tensor, but the renormalized semi-classical stress-energy tensor of a quantum field (SCSET) has a nonzero trace that depends on the local curvature of the background spacetime, but is independent of the particular quantum state of the field.[1] For a spin $s$ field in a 4D spacetime[2] the anomalous trace can be written as

$$T^\mu_\mu = \frac{m_p^2}{2^8 90\pi^2}\left[ a_s\left(R^{\alpha\beta\gamma\delta}R_{\alpha\beta\gamma\delta} - 2R^{\alpha\beta}R_{\alpha\beta} + \frac{1}{3}R^2 + \frac{2}{3}\Box R\right) + b_s\left(R^{\alpha\beta\gamma\delta}R_{\alpha\beta\gamma\delta} - 4R^{\alpha\beta}R_{\alpha\beta} + R^2\right) + c_s\Box R\right]. \quad (1.1)$$

Alternatively,

$$T^\mu_\mu = \frac{m_p^2}{2^8 90\pi^2}\left[a_s C^2 + b_s\left(C^2 - 2R^{\alpha\beta}R_{\alpha\beta} + \frac{2}{3}R^2 - \frac{2}{3}\Box R\right) + c'_s \Box R\right], \quad (1.2)$$

where $C^2 \equiv C^{\alpha\beta\gamma\delta}C_{\alpha\beta\gamma\delta}$ is the square of the Weyl tensor, $c'_s = c_s + 2(a_s + b_s)/3$, and $m_p$ is the Planck mass. My units are $G = c = 1$ and $\hbar = m_p^2$. An additional potential contribution to the trace from background gauge fields will not be considered.

In a Schwarzschild background the factor in front can be written in terms of the thermal pressure per helicity state of radiation at the Hawking temperature, $P_0 \equiv 2\sigma T_H^4 / 3$, as $P_0(2M)^4$. The coefficients $a_s$ and $b_s$, which times the overall numerical factor are often denoted $b$ and $b'$ in the literature, depend only on the spin of the field, The values are: for a spin 0 field $a_0 = 12$, $b_0 = -4$, $c_0 = 0$; for a spin 1/2 field $a_{1/2} = 72$, $b_{1/2} = -44$, and for a spin 1 field $a_1 = 144$, $b_1 = -248$, $c_1 = -240$. The values of $c_s$ depend on the renormalization scheme; the values given is for point-splitting renormalization. The quantity with the coefficient $b_s$ in Eq. (1.1) is the Euler density, whose integral is a topological invariant. For a Schwarzschild background, Eqs. (1.1) and (1.2) simplify to $T^\mu_\mu = q_s P_0 x^6$, with $x \equiv 2M/r$ and $q_s = 12(a_s + b_s)$.

The question addressed in this paper is how to identify an "anomalous part" $\langle T^{\mu\nu} \rangle_{anom}$ of the full SCSET. This "ASET" must have the trace of Eqs. (1.1)/(1.2) and, I argue, should be defined, like the trace, to be independent of the quantum state. There is no good reason to associate any part of the state-dependence of the full semi-classical stress-energy tensor (SCSET) with the ASET, and I will show that attempts to do so have not been very successful. An ASET can be derived by constructing an anomalous effective action $S_{anom}$ as a scalar functional of the metric. Varying with respect to the metric generates an ASET

$$\langle T_{\mu\nu} \rangle_{anom} = -\frac{2}{\sqrt{-g}} \frac{\delta}{\delta g^{\mu\nu}} S_{anom} \tag{1.3}$$

that is automatically conserved. In a 4D spacetime no such effective action exists as a *local* scalar functional of the metric constructed just from the curvature tensors and their derivatives, with the exception of the part of the anomalous effective action generating a $\Box R$ term in the trace.

I consider an ASET acceptable only if it satisfies certain physical regularity conditions. For a stationary black hole background, the ASET should be stationary and regular in a freely falling frame at the horizon. The Killing horizon of a stationary black hole is not a boundary of the spacetime. It is a regular null hypersurface. Also, for a black hole in an asymptotically flat spacetime the asymptotic behavior at infinity of the ASET should be consistent with the asymptotic flatness. One would expect the energy density should fall off rapidly enough (faster than $r^{-3}$) that the integral for the contribution to the total gravitational mass of the system converges. A slower, $r^{-2}$, falloff of a positive energy density and radial stress equal to an asymptotic positive energy flux can be associated with an outward flow of radially streaming positive energy radiation.

There is no plausible rationale as to why any such outflow should be state-independent and be considered part of the ASET. In the Hartle-Hawking state the asymptotic stress tensor of the full SCSET is isotropic at order $r^{-2}$, as one expects for thermal radiation in equilibrium in the gravitational field of the black hole, with no sign of any anisotropy associated with the trace anomaly. In the Unruh state there is an asymptotic energy density, radial stress, and energy flux falling as $r^{-2}$, but there is no reason to associate this with the trace anomaly in 4D. The energy flux of the Hawking radiation itself does not violate asymptotic flatness, since to be in the Unruh state the black hole must have formed at a finite time in the past. The radiation would only extend out to the intersection with the future light cone from the formation event and would contribute a finite amount to the total gravitational mass of the system.

Three prominent attempts to construct an ASET in the literature give quite discordant results, none of which are satisfactory according to the above criteria. I first consider in Part II the approach pioneered by Brown and Ottewill[3] (BO) and refined by Brown, Ottewill and Page[4] (BOP). It is based on an expression for the *change* $W\left(e^{-2\omega}g_{\mu\nu}\right) - W\left(g_{\mu\nu}\right)$ in the effective action under a conformal transformation of the metric, $\tilde{g}_{\mu\nu} = e^{-2\omega}g_{\mu\nu}$, with $g_{\mu\nu}$ the physical metric. The BO effective action is a functional of the metric and $\omega$. If the physical spacetime is Ricci-flat, with a static Killing vector field $K^{\alpha}$, choosing $\omega = \frac{1}{2}\log\left|g_{\alpha\beta}K^{\alpha}K^{\beta}\right|$, so the conformal spacetime is "ultrastatic" $\left(\tilde{g}_{00} = -1\right)$, the trace anomaly in the conformal spacetime is drastically simplified. It vanishes completely for spin 0. In this case, if one makes the (highly dubious) *assumption* that the full anomalous effective action vanishes in the ultrastatic spacetime, an expression for the anomaly effective action in the physical spacetime follows.

The problem is that this ASET is singular on the horizon, since $\omega$ is singular on the horizon. While the singular behavior in the full SCSET can be canceled by adding conformally invariant thermal terms (also based on the static Killing vector) to the effective action, the thermal terms are only appropriate for the Hartle-Hawking[5] state. There was no rationale provided for dealing with other quantum states, such as the physically more interesting Unruh state. For spin 0 the Brown-Ottewill approach does result in the fairly good Page approximation[6] to the SCSET, but an extension to the spin 1 Hartle-Hawking state attempted in BOP did not lead to satisfactory results.

An explicitly nonlocal anomaly effective action in 4D, similar in spirit to the 2D Polyakov action[7], was introduced by Riegert[8]. The action can be made local by introducing two auxiliary scalar fields satisfying fourth-order wave equations sourced by parts of the trace anomaly. The asymptotic behavior of the corresponding ASET on the horizon and at infinity for conformally coupled quantum fields in the Schwarzschild background was worked out by Balbinot, et al[9] (BFS). An equivalent anomaly effective action, based a somewhat different definition of the auxiliary scalars, was analyzed in greater detail by Mottola and Vaulin[10] (MOTV), who tried to argue that there was enough freedom in the solutions for the auxiliary scalars that the corresponding ASET could approximate the full range of the SCSET for multiple quantum states. However, their construction of the ASET for the Schwarzschild background contained sign errors, and the actual ASET, as recognized by BFS, had unacceptable asymptotic behavior at infinity. BFS suggested modifying the coefficient of a conformally invariant term to the action in order to improve the asymptotic behavior, with the goal of making the asymptotic behavior consistent with the Unruh state.

An alternative anomalous effective action introduced recently by Mottola[11] (MOT2) modifies the Riegert action to require only one auxiliary scalar field. While the MOT2 paper did not consider black holes, and is focused on the contribution to the anomaly from non-gravitational gauge fields, it is straightforward to use the MOT2 effective action to calculate the ASET for the Schwarzschild background. I show that asymptotic behavior of the MOT2 ASET is even more unphysical than that of the BFS/MOTV ASET. The detailed discussion of the BFS/MOTV and MOT2 actions and ASETs is in Part III.

The main result of this paper, presented in Part IV, is to show that within the general framework of the BFS/MOTV effective actions, with two auxiliary scalar fields, there is a unique ASET that is regular on the horizon, has the correct trace, and has an asymptotic $r^{-4}$ falloff of the energy density at large radii, consistent with asymptotic flatness. The ASET components are simple polynomials in $2M/r$ and, except for an overall factor, are the same for all spins. The action can be considered a linear combination of the MOTV and MOT2 effective actions. The ASET is state independent, and includes neither the Hawking radiation of the Unruh state nor the asymptotic thermal heat bath of the Hartle-Hawking state.

Numerical calculations of the full SCSET have been carried out for the spin 0 Hartle-Hawking quantum state in a Schwarzschild background by Howard[12] and later with higher action accurately by Anderson, et al[13], for the spin 1 Hartle-Hawking state by Jensen and Ottewill[14] (JO), for the spin 0 and spin1 Unruh[15] states by Jensen, et al[16] (JMO), and for the spin 0 Boulware state.[17]. While there is a close connection between the ASET and the SCSET in a 2D Schwarzschild background[18], this is just because in 2D the trace and energy-momentum conservation completely determine the SCSET within a couple of constants of integration, this is not the case in 4D, as we discuss briefly.

I make no attempt to discuss the part of the ASET associated with the □R term in the trace anomaly of Eq. (1.2), since this can be generated by a completely local effective action proportional to $\int R^2 \sqrt{-g} d^4 x$ and is not considered part of the "true" trace anomaly by MOTV and MOT2.

## II. THE BROWN-OTTEWILL EFFECTIVE ACTION AND ASET

My discussion of Brown-Ottewill effective action is based on its most mature development in the BOP paper. The form of the ASET js

$$\left\langle T_\nu^\mu \right\rangle_{\text{anom}} = a_s T_{A\nu}^\mu + b_s T_{B\nu}^\mu + c_s T_{C\nu}^\mu, \tag{2.1}$$

with $a_s$, $b_s$, and $c_s$ identical to those of Eq. (1.1). The thermal contributions to the total SCSET $\bar{T}^{\mu\nu}$ are tensors $T_{2\mu}^{\nu}$ and $T_{4\mu}^{\nu}$ constructed from the static Killing vector, and modify the terms in Eq. (2.1) to

$$\bar{T}_A^{\mu\nu} = T_A^{\mu\nu} + \frac{1}{3}T_2^{\mu\nu} + 2T_4^{\mu\nu}, \qquad (2.2)$$

$$\bar{T}_B^{\mu\nu} = T_B^{\mu\nu} + T_2^{\mu\nu} + 2T_4^{\mu\nu}, \qquad (2.3)$$

$$\bar{T}_C^{\mu\nu} = T_C^{\mu\nu} - \frac{1}{2}T_2^{\mu\nu} - T_4^{\mu\nu}. \qquad (2.4)$$

Explicit expressions for $T_2^{\mu\nu}$ and $T_4^{\mu\nu}$ in a Schwarzschild background are found in Eqs. (4.22) and (4.23) and for $\bar{T}_A^{\mu\nu}$, $\bar{T}_B^{\mu\nu}$, and $\bar{T}_C^{\mu\nu}$ in Eqs. (4.19)-(4.21) of BOP. Subtracting the appropriate linear combinations of the thermal contributions from the SCSET gives for the ASET transverse stresses

$$T_{A\theta}^{\theta} = \frac{P_0}{24(1-x)^2}\left(-64x^6 + 120x^7 - 57x^8\right), \qquad (2.5)$$

$$T_{B\theta}^{\theta} = \frac{P_0}{8(1-x)^2}\left(-32x^6 + 72x^7 - 39x^8\right), \qquad (2.6)$$

$$T_{C\theta}^{\theta} = \frac{P_0}{16(1-x)^2}\left(128x^6 - 264x^7 + 135x^8\right). \qquad (2.7)$$

The ASET radial stresses are

$$T_{Ar}^{r} = \frac{P_0}{24(1-x)^2}\left(32x^6 - 96x^7 + 63x^8\right), \qquad (2.8)$$

$$T_{Br}^{r} = \frac{P_0}{8(1-x)^2}\left(16x^6 - 48x^7 + 33x^8\right), \qquad (2.9)$$

$$T_{Cr}^{r} = \frac{P_0}{16(1-x)^2}\left(-64x^6 + 144x^7 - 81x^8\right). \qquad (2.10)$$

I use the notation $x \equiv 2M/r$ in place of the BOP $\omega$. The $T_t^t$ components follow from the traces, which for spin 0 are $T_{A\mu}^{\mu} = T_{B\mu}^{\mu} = 12P_0 x^6$ and $T_{C\mu}^{\mu} = 0$.

The thermal contributions introduced by BOP are only appropriate for the Hartle-Hawking state, and BOP give no indication for how their approach can be modified for the Unruh state SCSET. The fact that $T_{C\nu}^{\mu}$ is nonzero even though its trace vanishes in a Ricci-flat geometry is an example of why the key assumption behind the entire derivation is suspect. The approximation to the Hartle-Hawking SCSET reduces to the Page approximation for spin 0. However, for spin 1 the BOP SCSET has an energy density on the horizon that is about a factor of 5 smaller in magnitude than the numerical value calculated by JO. The divergence of BOP ASET at the horizon, means that it cannot be taken seriously as any kind of state-independent quantity, and cannot fulfill what I consider the proper role of an ASET.

## III. THE BFS/MOTV and MOT2 EFFECTIVE ACTIONS AND ASETS

I start from the Riegert-based effective action in the form given in BFS, with somewhat modified notation. There are two auxiliary scalar fields $\bar\phi$ and $\bar\psi$ sourced by the trace anomaly. Let

$$A \equiv C^2, \quad B \equiv {}^*R_{\alpha\beta\gamma\delta}{}^*R^{\alpha\beta\gamma\delta} - \frac{2}{3}\Box R. \tag{3.1}$$

The anomalous effective action from Eq. (3.6) of BFS associated with the first two terms of Eq. (1.2) is

$$S_{\rm BFS} = m_{\rm p}^2 \int d^4x \sqrt{-g} \left[ \begin{array}{c} \frac{1}{2}\bar\phi\,\Delta_4\bar\phi - \dfrac{1}{8\pi\sqrt{-1440 b_s}}\left[a_s A + b_s B\right]\bar\phi \\[6pt] -\dfrac{1}{2}\bar\psi\,\Delta_4\bar\psi + \dfrac{a_s}{8\pi\sqrt{-1440 b_s}}\gamma A\bar\psi \end{array} \right]. \tag{3.2}$$

The third term in Eq. (1.2) is put aside here, since it can be generated by a purely local effective action. The parameter $\gamma$ is the only nontrivial modification of the coefficients in the Riegert action that preserves the trace anomaly. It was introduced as a fudge factor by BFS because the original $\gamma = 1$, led to physically unacceptable results for a Schwarzschild background. The operator $\Delta_4$ is the unique conformally invariant fourth order scalar operator in four dimensions,

$$\Delta_4 = \Box^2 + 2R^{\alpha\beta}\nabla_\alpha\nabla_\beta - \frac{2}{3}R\Box + \frac{1}{3}(\nabla^\alpha R)\nabla_\alpha. \tag{3.3}$$

Eq. (3.2) can be simplified by rescaling the auxiliary scalar fields. Let

$$\bar\phi = -\frac{\sqrt{-b_s}}{4\pi\sqrt{1440}}\chi, \quad \bar\psi = \frac{a_s}{4\pi\sqrt{-1440 b_s}}\psi. \tag{3.4}$$

Then

$$S_{\rm BFS} = \frac{m_{\rm p}^2}{90\pi^2 2^8} \int d^4x \sqrt{-g}\left[ -\frac{1}{2}b_s \chi\Delta_4\chi + \frac{1}{2}(a_s A + b_s B)\chi + \frac{1}{2}\frac{a_s^2}{b_s}(\psi\Delta_4\psi - \gamma A\psi)\right]. \tag{3.5}$$

Varying the action with respect to the auxiliary scalars gives the equations

$$\Delta_4 \chi = \frac{1}{2}\left(B + \frac{a_s}{b_s}A\right), \quad \Delta_4 \psi = \frac{1}{2}\gamma A. \tag{3.6}$$

The MOTV action is in fact completely equivalent to Eq. (3.5), with $\gamma = 1$. Substitute

$$\chi \to \varphi - \frac{a_s}{b_s}\psi. \tag{3.7}$$

With $\gamma = 1$ the action splits into two pieces associated separately with each of the first two terms of the trace anomaly,

$$S_{\rm MV} = \frac{m_{\rm p}^2}{90\pi^2 2^8}(a_s S_A + b_s S_B), \tag{3.8}$$

$$S_A = \frac{1}{2}\int d^4x\sqrt{-g}\left(-\varphi\Delta_4\psi - \psi\Delta_4\varphi + A\varphi + B\psi\right), \tag{3.9}$$

$$S_B = \frac{1}{2}\int d^4x\sqrt{-g}\left(-\varphi\Delta_4\varphi + B\varphi\right). \tag{3.10}$$

Varying with respect to the metric gives the ASET in terms of two separately conserved tensors,

$$\langle T_{\mu\nu}\rangle_{\text{anom}} = \frac{m_p^2}{90\pi^2 2^8}\left(a_s A_{\mu\nu}(\varphi,\psi) + b_s B_{\mu\nu}(\varphi)\right), \tag{3.11}$$

with $A_\mu^\mu = A$ and $B_\mu^\mu = B$. The equation for $\varphi$ is just $\Delta_4\varphi = (1/2)B$.

Simplifying to a Ricci-flat background, in which $\Delta_4 = \Box^2$, the explicit expressions for $A_{\mu\nu}$ and $B_{\mu\nu}$ are, from Eqs. (3.42) and (3.41) of MOTV,

$$\begin{aligned}A_{\mu\nu} &= -2(\nabla_{(\mu}\varphi)(\nabla_{\nu)}\Box\psi) - 2(\nabla_{(\mu}\psi)(\nabla_{\nu)}\Box\varphi) \\ &\quad +2\nabla^\alpha\left[(\nabla_\alpha\varphi)(\nabla_\mu\nabla_\nu\psi)+(\nabla_\alpha\psi)(\nabla_\mu\nabla_\nu\varphi)\right] - g_{\mu\nu}(\Box\varphi)(\Box\psi) \\ &\quad -\frac{4}{3}\nabla_\mu\nabla_\nu\left[(\nabla_\alpha\varphi)(\nabla^\alpha\psi)\right] + \frac{1}{3}g_{\mu\nu}\Box\left[(\nabla_\alpha\varphi)(\nabla^\alpha\psi)\right] \\ &\quad -4C_{\mu\alpha\nu\beta}\left(\nabla^\alpha\nabla^\beta\varphi + \nabla^\alpha\nabla^\beta\psi\right) + \frac{2}{3}g_{\mu\nu}\Box^2\psi - \frac{2}{3}\nabla_\mu\nabla_\nu\Box\psi,\end{aligned} \tag{3.12}$$

and

$$\begin{aligned}B_{\mu\nu} &= -2(\nabla_{(\mu}\varphi)(\nabla_{\nu)}\Box\varphi) + 2\nabla^\alpha\left[(\nabla_\alpha\varphi)(\nabla_\mu\nabla_\nu\varphi)\right] - \frac{1}{2}g_{\mu\nu}(\Box\varphi)^2 \\ &\quad -\frac{2}{3}\nabla_\mu\nabla_\nu\left[(\nabla_\alpha\varphi)(\nabla^\alpha\varphi)\right] + \frac{1}{6}g_{\mu\nu}\Box\left[(\nabla_\alpha\varphi)(\nabla^\alpha\varphi)\right] \\ &\quad -4C_{\mu\alpha\nu\beta}\nabla^\alpha\nabla^\beta\varphi + \frac{2}{3}g_{\mu\nu}\Box^2\varphi - \frac{2}{3}\nabla_\mu\nabla_\nu\Box\varphi.\end{aligned} \tag{3.13}$$

In any Ricci-flat background $B = C^2$, and the equations for $\varphi$ and $\psi$ are identical.

Further simplifying to the static, spherically symmetric Schwarzschild background within the MOTV framework, the $\varphi$ and $\psi$ consistent with a stationary, axisymmetric ASET are

$$\varphi(r,t) = \varphi(r) + p\frac{t}{2M}, \quad \psi(r,t) = \psi(r) + p'\frac{t}{2M}. \tag{3.14}$$

Nonzero parameters $p$ and/or $p'$ allow for a nonzero energy flux in the ASET, without making the ASET time dependent. The solutions for $\varphi(r)$ and $\psi(r)$ can differ in the constants of integration. Only three of these for each field are non-trivial, since only derivatives of the scalars appear in Eqs. (3.12) and (3.13). Using $x = 2M/r$ as the radial variable, the general solution has

$$-2M\frac{\partial\varphi}{\partial r} = x^2\frac{\partial\varphi}{\partial x} = \frac{2}{3} + x - \frac{c_H x^2}{1-x} + \frac{q}{6}\left[1 + 3x - \frac{2x^2}{1-x}\log(x)\right]$$
$$- \left(x^{-1} + 1 + x\right)\left[c_\infty + \frac{1}{3}(q-2)\log(1-x)\right]. \tag{3.15}$$

Denote the corresponding constants of integration in $\psi$ by $q'$, $c'_H$, and $c'_\infty$. The explicit expressions MOTV give for $A^\nu_\mu$ and $B^\nu_\mu$ ($F^\nu_\mu$ and $E^\nu_\mu$ in their notation) in their Appendix A have the wrong signs for the terms in the diagonal components containing $p^2$ and $pp'$. The problem is that they use the wrong (Euclidean) sign for $g^{tt}$ in evaluating $\nabla_\alpha \varphi \nabla^\alpha \varphi$ and $\nabla_\alpha \varphi \nabla^\alpha \psi$.

MOTV, as did BFS, try to adapt the ASET to different quantum states by choosing different constants of integration in $\varphi$ and $\psi$ for the Hartle-Hawking, Unruh, and Boulware quantum states in the Schwarzschild background, and claim (falsely) some success in approximating the numerical results for the *full* SCSET with just the ASET. However, their "fits" with $p \neq 0$, as in their Hartle-Hawking and Unruh state models, are invalid because of the sign errors. (Their Boulware state model has $p = p' = 0$ and thus is not affected by the sign errors, but the Boulware state is not a valid quantum state for a black hole.) To have a nonzero energy flux, as in the Unruh state, $p$ or $p'$ must be non-zero, with

$$-(2M)^4 A^r_t = 2(p+p')x^2, \quad -(2M)^4 B^r_t = 2px^2, \tag{3.16}$$

so

$$-\langle T^r_t \rangle^{\text{anom}} = 2fP_0 = 2\left[(p+p')a_s + pb_s\right]P_0. \tag{3.17}$$

I have not tried to implement the MOTV program with the sign errors corrected, because as argued above I think the attempt is misguided, and would not be successful. The sources for the scalars are state-independent, and boundary conditions should just follow from regularity conditions at the horizon (for a black hole) and at infinity. If the regularity conditions are applied to $A^\nu_\mu$ and $B^\nu_\mu$ individually, the constants of integration are uniquely determined. There is no reason why the full SCSET should be well approximated by just the ASET.

Regularity at the horizon, $x=1$, requires $q = q' = 2$ to eliminate the $\ln(1-x)$ from Eq. (3.15). Also, singular $(1-x)^{-2}$ terms in the diagonal components of the full ASET at the horizon are eliminated if and only if

$$2a_s(c_H c'_H + pp') + b_s(c_H^2 + p^2) = 0. \tag{3.18}$$

This differs from the condition stated in Eq. (5.14a) of MOTV because of their sign errors.

The issue of terms behaving as $(1-x)^{-1}$ at the horizon is more complicated. Energy conservation for a stationary ASET requires that $\partial_r(r^2 T^r_t) = 0$. The static frame energy flux $F = -T^r_t/(1-x)$ and if not zero everywhere is singular on the horizon, where in the limit a static observer sees any ingoing radiation as infinitely blueshifted, and any outgoing radiation as infinitely redshifted. Physical regularity corresponds to regularity in a free fall frame, or $E^{\text{reg}} = E + F$ and $P_r^{\text{reg}} = P_r + F$ finite at the horizon, with $E^{\text{reg}} + P_r^{\text{reg}}$ vanishing proportional to $1-x$. The positive energy flux of the Unruh state Hawking radiation requires an infinitely negative $E$ and $P_r$ at horizon, corresponding to an infinitely blueshifted inflow of negative energy. Applied to the ASET, this additional regularity conditions becomes

$$f = a_s(p + p') + b_s p = a_s(c_H + c_H') + b_s c_H. \tag{3.19}$$

If the ASET energy flux vanishes, as it should for a state-independent ASET, the simple way to ensure regularity on the horizon is to take $c_H = c_H' = p = p' = 0$ in addition to $q = q' = 2$. Then the $A_\mu^\nu$ and $B_\mu^\nu$ tensors are each regular on the horizon, rather than just the combination that gives the ASET.

The asymptotic behavior at large $r$, $x \to 0$, is given, for $q = q' = 2$, by

$$-2M \frac{\partial \varphi}{\partial r} \to 1 - c_\infty(x^{-1} + 1) + O(x) \tag{3.20}$$

and correspondingly for $\partial \psi / \partial r$. To achieve the most rapid possible falloff in the asymptotic energy density, $E = -T_t^t$, I set $c_\infty' = c_\infty = 0$. Then the higher radial derivatives of the scalars fall off as $x^2$ or faster. The term in $A_t^t$ that dominates asymptotically is $-\Box \varphi \Box \psi$, with

$$(2M)^2 \Box \varphi = (2M)^2 \Box \psi \to 2x(2M) \frac{\partial \varphi}{\partial r} \to -2x, \tag{3.21}$$

and in $B_t^t$ is $-(\Box \varphi)^2 / 2$. The asymptotic energy density is then

$$E \to 2(2a_s + b_s) x^2 P_0. \tag{3.22}$$

The full asymptotic ASET has radial stress $P_r \to E$, and transverse stress $P_t = O(x^3)$. The asymptotic energy density is positive, since $2a_s + b_s > 0$ for all spins.

The asymptotic behavior of the energy density can be interpreted as radially propagating radiation, which is purely outgoing only if the ASET has a matching energy flux with $f = 2a_s + b_s$, most simply accomplished by setting $p = p' = 1$ There is no apparent source for any ingoing component of the radiation for an isolated black hole. A contribution to the gravitational mass of the system $\Delta M = \int 4\pi r^2 E \, dr$ increasing linearly with radius can then be understood as the accumulation of previously emitted radiation, terminating at the intersection with the future light cone from the black hole formation event. The problem is that the energy flux consistent with the asymptotic $E$ and $P_r$ implies a luminosity substantially larger than the Hawking luminosity.

For both spin 0 and spin 1, $2a_s + b_s = 20h_s$, where $h_s$ is the number of helicity states for spin $s$. The usual parameterization of the Hawking luminosity is

$$L_{\rm H} = \frac{4\pi}{245960\pi^2} \frac{m_{\rm p}^2}{M^2} k_s = 6\pi M^2 P_0 k_s, \tag{3.23}$$

with $k_1 = 6.4928$ and $k_2 = 0.7404$ from calculations of Page[19] and $k_0 = 14.36$ from Elster[20] and Taylor, et al.[21] The ratio of the luminosity associated with the BFS/MOTV ASET to the Hawking luminosity would be

$$L_{\rm MV}/L_{\rm H} = (320 h_s / 3 k_s), \tag{3.24}$$

7.43 for spin 0 and 32.9 for spin 1.

The BFS/MOTV anomalous effective action with $\gamma = 1$ is thus incompatible with the standard results for the Hawking radiation and the full SCSET appropriate to the Unruh state. This difficulty was noted by BFS, and motivated them to include the fudge factor $\gamma$ in Eq. (3.2). The term with the fudge factor is conformally invariant, since the Weyl tensor is conformally invariant, so this modification has no effect on the trace of the ASET. The fudge factor can be tuned to make the asymptotic energy density agree with the Hawking energy flux. However, the only satisfactory state-independent modification of the ASET is to make the coefficient of $x^2$ in the asymptotic energy density vanish, as discussed in Part IV.

Now consider the anomalous effective action proposed in MOT2, based on a single scalar field $\chi$. This action is

$$S_{\rm M} = \frac{m_{\rm p}^2}{90\pi^2 2^8} S_D, \quad S_D = \frac{1}{2}\int d^4x \sqrt{-g}\left[-b_s \chi \Delta_4 \chi + (b_s B + a_s A)\chi\right], \tag{3.25}$$

with $A$ and $B$ from Eq. (3.1). This action can be obtained from Eq. (3.5) by setting $\gamma = 0$ to eliminate the last term completely. Then the equation for $\psi$ becomes $\Delta_4 \psi = 0$, with the trivial solution $\psi = 0$, for which the action reduces to Eq. (3.25). Assuming Ricci flatness of the background, the expression for the MOT2 ASET is

$$\langle T_{\mu\nu} \rangle_{\rm anom} = \frac{m_{\rm p}^2}{90\pi^2 2^8}\left[b_s B_{\mu\nu}(\chi) - 4\pi a_s C_{(\mu\ \nu)}^{\ \alpha\ \beta} \nabla_\alpha \chi \nabla_\beta \chi\right], \tag{3.26}$$

with $\chi$ substituted for $\varphi$ in Eq. (3.13). Also, the equation for $\chi$ becomes

$$\Box^2 \chi = \frac{1}{2}\left(1 + \frac{a_s}{b_s}\right) C^2, \tag{3.27}$$

so a solution for $\chi$ is just the factor $(1 + a_s/b_s)$ times a solution for $\varphi$ or $\psi$ in the MOTV ASET. For a Schwarzschild black hole, since there is only one auxiliary scalar field, the analogue of Eq. (3.18) *requires* $c_{\rm H} = p = 0$. A vanishing asymptotic energy density requires $c_\infty = 0$. The unique acceptable solution for $\chi$ has

$$-2M \frac{\partial \chi}{\partial r} = x^2 \frac{\partial \chi}{\partial x} = \left(1 + \frac{a_s}{b_s}\right)\left[1 + x - \frac{2}{3}\frac{x^2}{1-x}\log(x)\right]. \tag{3.28}$$

The asymptotic energy density and radial stress are then

$$E, P_{\rm r} \to 2\frac{(a_s + b_s)^2}{b_s} P_0 x^2, \tag{3.29}$$

which are *negative*, since for all spins $b_s < 0$. This makes absolutely no physical sense in terms of radiation. There can be no net energy flux with $p = 0$, so Eq. (3.29) corresponds to equal amounts of incoming and outgoing negative energy radiation. I conclude that the MOT2 ASET is even less viable than the MOTV ASET for a Schwarzschild black hole.

## IV. A PHYSICALLY ACCEPTABLE ASET

To be physically acceptable the ASET in an asymptotically flat spacetime should have an energy density falling off more rapidly than $x^3$, so that the contribution of the ASET to the total gravitational mass of the system is finite. This can be accomplished by the correct choice of the fudge factor $\gamma$. The asymptotic energy density is determined by the terms in the action quadratic in the auxiliary scalars. For a Schwarzschild background, the contribution from the terms quadratic in $\chi$ is given by Eq. (3.29), and together with the term quadratic in $\psi$, since $\Delta_4 \psi = \gamma C^2 / 2$,

$$E = -T_t^t \rightarrow \left[ \frac{(a_s + b_s)^2}{b_s} - \gamma^2 \frac{a_s^2}{b_s} \right] 2x^2 P_0. \tag{4.1}$$

The choice

$$\gamma = \pm \frac{(a_s + b_s)}{a_s} \tag{4.2}$$

cancels the asymptotic $x^2$ falloffs and in fact all contributions to the ASET quadratic in the auxiliary scalars. The terms linear in the scalars arising from variations of $C^2$ also cancel. The ASET does not depend on the sign of $\gamma$. The ASET reduces to the contribution from the $\Box R$ term in $B$ in Eq. (3.5). As can be seen from the contribution of a similar term in the MOTV $S_B$ to $B_{\mu\nu}$, this is

$$\langle T_{\mu\nu} \rangle_{\text{anom}} = \frac{2}{3} b_s P_0 (2M)^4 \left[ -\nabla_\mu \nabla_\nu \Box \chi + g_{\mu\nu} \Box^2 \chi \right]. \tag{4.3}$$

This version of the ASET in the Schwarzschild black hole background, with constants of integration in the auxiliary scalars fixed by regularity on the horizon and infinity, so $\chi$ is given by Eq. (3.28), is remarkably simple. All components of the ASET are polynomials in $x = 2M/r$, since the logarithm in Eq. (3.28) is not present in $\Box \chi$. Explicitly, for any spin,

$$\langle T_t^t \rangle_{\text{anom}} = -E = -(a_s + b_s) \frac{2}{3} x^4 \left(1 + x - 5x^2\right) P_0, \tag{4.4}$$

$$\langle T_r^r \rangle_{\text{anom}} = P_r = (a_s + b_s) \frac{2}{3} x^3 \left(4 + x + x^2 - 3x^3\right) P_0, \tag{4.5}$$

$$\langle T_\theta^\theta \rangle_{\text{anom}} = P_t = -(a_s + b_s) \frac{4}{3} x^3 \left(1 - 4x^3\right) P_0. \tag{4.6}$$

For spin 0, $a_s + b_s = 8$, and for spin 1, $a_s + b_s = -104$. At large $r$ the ASET is dominated by the stresses falling as $r^{-3}$, with the energy density falling off more rapidly, as $r^{-4}$. The components for spin 0 are plotted in Fig. 1.

I consider the ASET of Eq. (4.3) the only physically viable ASET for a Schwarzschild black hole within the framework established by BFS. Of course, there are potentially other ways of adding conformally invariant terms to the Riegert effective action beyond what BFS considered.

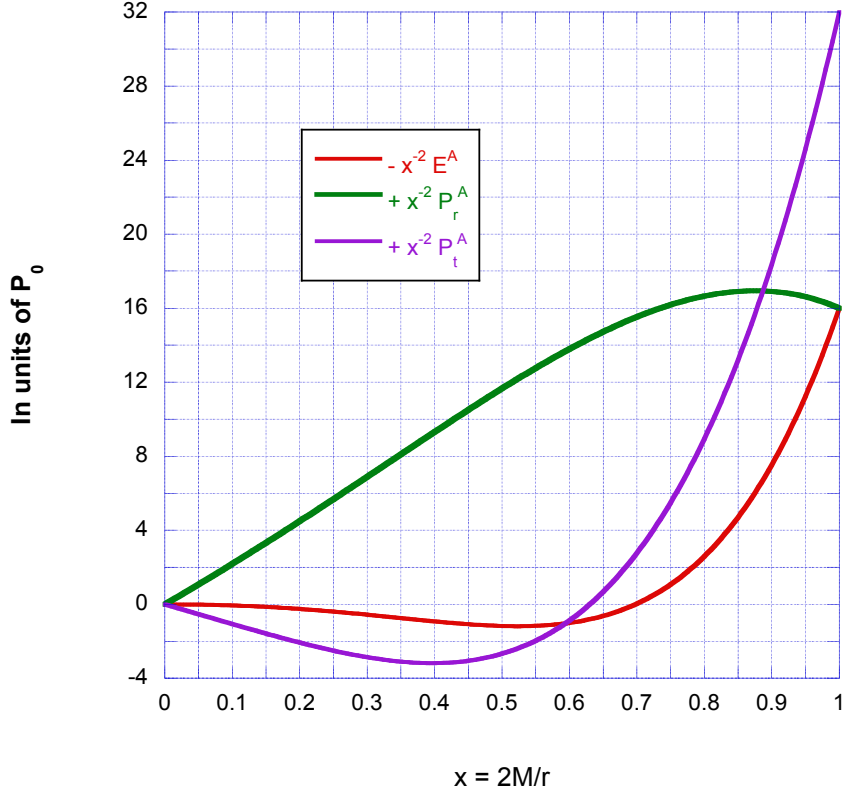

Figure 1. The static frame components of the spin 0 ASET from Eqs. (4.4)-(4.6), divided by $x^2$, are shown. To get the values of the components for spin 1, multiply each by $-13$.

Is there any sense in which one might consider this ASET an approximation to the full SCSET? One might hope so. Unfortunately, the answer is no. For the spin 0 SCSET, fits to numerical data calculated by Anderson, et al and JMO, as tabulated by Visser[22], give for the transverse stress of the Hartle-Hawking state

$$P_t^{HH} = \left(1 + 2x + 3x^2 + 3.65x^3 + 14.40x^4 - 48.17x^5 + 34.39x^6\right)P_0 \quad (4.7)$$

and the Unruh state

$$P_t^U = \left(0.253x^3 + 25.544x^4 - 57.666x^5 + 37.617x^6\right)P_0. \quad (4.8)$$

Neither has any resemblance to the $P_t$ of the ASET in Eq. (4.6), and the values at the horizon are considerably smaller. The spin 0 $E$ and $P_r$ of the SCSET have substantial state-dependent contributions, so there is no expectation that they should resemble the ASET.

For spin 1 the contribution of the Hawking radiation to the Unruh state SCSET is quite insignificant except *very* close to $x=0$ and $x=1$. Visser[23] was able to provide a computer file of the JMO numerical data. A polynomial fit to the data for $P_t$, reasonably accurate for $x > 0.4$, but with a somewhat tentative extrapolation to $x=0$, is[24]

$$P_t = 2P_0\left(40.9x^3 - 385.2x^4 + 32.7x^5 - 471.1x^6\right). \tag{4.9}$$

The corresponding radial stress implied by momentum conservation and the trace anomaly, including the constraint of consistency with the Hawking flux, is

$$P_r = 2P_0\left[2.435x^2\left(1 - \frac{0.5}{1-x}\right) - 79.4x^3 + 367.2x^4 - 39.8x^5 + 220.25x^6\right]. \tag{4.10}$$

These are compared with the corresponding stresses of the ASET in Fig. 2. The term in Eq. (4.10) singular at $x=1$ is associated with infinite blueshift in the static frame of the negative energy inflow into the black hole required by presence of Hawking radiation.

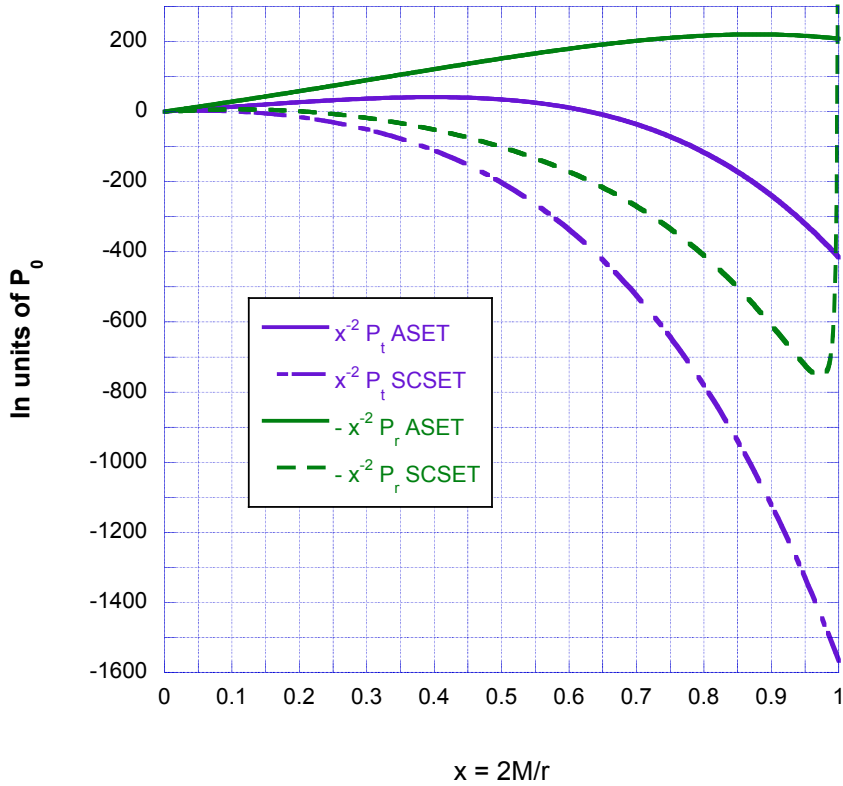

Figure 2. The spin 1 ASET stresses are compared with the Unruh state SCSET stresses.

The trace anomaly at the horizon is $T_\mu^\mu(1) = -1248 P_0$. Approaching, but not very close to the horizon, the SCSET radial stress has the opposite sign and is much larger in magnitude than that of the ASET. This is the simple result of the sign of $2P_t - T_\mu^\mu$. Other than having the same trace, it is hard to say that the ASET gives any insight into the behavior of the SCSET. For more detail on the properties of the SCSET see Ref. 24. By the way, the BFS attempt to match the asymptotic energy density with the Hawking energy flux in the ASET requires, for spin 1, $\gamma^2 = 0.5362$, very close to my value of $\gamma^2 = 0.5216$. The BFS Unruh state model would be just as poor a match with the Unruh state numerical results as my ASET.

What has been ignored in all this is that by far the dominant contribution to the overall SCSET is expected to be the contribution from spin 2 gravitons. The spin 2 trace anomaly in the Schwarzschild background is $20352 P_0$, more than 10 times the magnitude of the spin 1 trace anomaly[25]. This sets the scale of the SCSET, even though the spin 2 Hawking luminosity is almost 10 times smaller than the spin 1 Hawking luminosity. It is also possible that the full quantum gravity contribution to the SCSET is qualitatively different from that of non-gravitational fields.

## V. FINAL THOUGHTS

I have analyzed three prominent proposals for anomalous effective actions generating the trace anomaly of the stress-energy tensor in curved spacetimes, and in the context of the Schwarzschild black hole background argue that none of them should be considered physically acceptable. The ASET from the Brown-Ottewill effective action, without the clearly state dependent thermal terms, is singular on the horizon. The Riegert nonlocal effective action, equivalent to the MOTV effective action, generates an ASET with a physically unacceptable asymptotic behavior at large radii, as does the MOT2 effective action. BFS showed that a simple modification of the Riegert action, equivalent to adding a conformally invariant term, can be adjusted to make the asymptotic behavior consistent with the asymptotic behavior of the SCSET for the Unruh state, but this kind of state-specific tampering seems inconsistent with what should be expected of an ASET.

The main result of this paper is the demonstration that the coefficient of the BFS modification of the Riegert effective action can be chosen to give an ASET well behaved both on the black hole horizon and asymptotically, with an energy density falling as $r^{-4}$. This ASET is not in conflict with the SCSET for any stationary quantum state consistent with the actual existence of a black hole horizon, but it does not constitute a reasonable approximation to the SCSET for any quantum state. in 4D. Unlike in the 2D version of the Schwarzschild background[18], the trace anomaly has essentially no connection with the existence or magnitude of the Hawking luminosity of a black hole.

Whether this is the final word on the subject is unclear. For both spin 0 and spin 1 fields this ASET does not seem to be very helpful in understanding the numerical results for the expectation value of the full renormalized semi-classical stress-energy tensor. If one did have confidence in a particular ASET, it might be of some help in understanding the semi-classical backreaction on the geometry in the interior of the black hole, which is outside the scope of all numerical calculations to date. Of course, the semi-classical approximation will definitely break down in the deep interior of the black hole, as the spacetime curvature approaches the Planck scale.

An anomaly effective action that gives rise to obviously unphysical results for a Schwarzschild background may not do so for certain other backgrounds. Still, I consider this a fatal flaw. The effective action should be valid for any background, though the boundary conditions on the auxiliary scalars may be background dependent. My proposal for an anomaly effective action should be tested in a broader context and should not be considered any kind of final answer.

To date there are no calculations of the SCSET for the quantum gravitational field. Until this is rectified, there can be no claim of understanding of the overall semi-classical corrections to the geometry of black holes.

ACKNOWLEDGEMENTS

I thank Andreas Karch and Hal Haggard for reading earlier versions of the manuscript and offering helpful comments. This research was supported in part by Perimeter Institute for Theoretical Physics. Research at Perimeter Institute is supported by the Government of Canada through the Department of Innovation, Science, and Economic Development, and by the Province of Ontario through the Ministry of Research and Innovation.